\documentstyle[preprint,aps]{revtex}
\begin{document}
\draft
\title {THE ISENTROPIC EQUILIBRIUM DECONFINING PHASE TRANSITION IN
        THE EXTENDED BAG MODEL}
 
 \author{V. Gogohia, B. Luk\'acs and M. Priszny\'ak}
 
\address{RMKI, Department of Theoretical Physics,
         Central Research Institute for Physics,  \\
         H-1525,  Budapest 114,  P. O. B.  49, Hungary }
 
\maketitle
 
\begin{abstract}
By using an Ansatz for the pressure (measured in terms
of the bag constant) of the hadronic gas in equilibrium, we have
formulated a new bag-type model for the phase transition, i. e.
the extended bag model. This allows one
to take into account the nonperturbative vacuum effects from both
sides of the equilibrium condition. We have explicitly shown that
our model automatically provides an isentropic equilibrium
deconfining phase transition from the quark-gluon plasma
phase to the hadronic gas phase with a temperature and chemical
potential dependent bag pressure at the phase boundary.
This makes it possible to regulate
the total entropy and baryon number in both phases, i. e. the flow
of the specific entropy per baryon across the phase boundary.
We have also determined the functional form of the temperature and
chemical potential dependence of the bag pressure outside the
phase boundary.
\end{abstract}

\pacs{PACS numbers: 05.70.Fh, 12.38.Aw, 12.38.Lg, 12.38 Mh}

\vfill
\eject

   I. The quark-gluon plasma (QGP) phase is a necessary
step in the evolution of the Exited Matter from Big Bang to the
present days. Apparently, the only way to study this phase
of the expansion of the Universe is nuclear (heavy ion) collisions
at high energies which makes it possible "to recreate conditions
akin to the first moments of the Early Universe, the Big Bang, in
the laboratory" [1]. Because of the confinement phenomenon,
the nonperturbative vacuum structure must play a very
important role in the transition from QGP to the formation of the
hadronic particles (i. e., hadronization) and vice versa. As it
was underlined in our papers [2], any correct model of the
nonperturbative effects such as quark confinement or dynamical
chiral symmetry breaking (DCSB) becomes a model of the true QCD
ground state (i. e., the nonperturbative vacuum) and the other way
around. Thus the difference between the perturbative (which
is always
normalizable to zero) and the nonperturbative vacua appears to
be necessarily nonzero and finite so that it describes the above
mentioned nonperturbative phenomena at zero temperature. The
existence of the finite vacuum energy per unit volume - the bag
constant [3, 4] - becomes important for a realistic calculation
of the transition from hadronic to QGP phases and vice versa
at nonzero temperature as well.
 
There are two main approaches to investigate QGP,
namely the resummed finite-temperature perturbation theory
(effective field theory method) [5-7] and the lattice one
[8]. The former breaks down after the fifth
order in the QCD coupling constant $g$ because of the severe
infrared divergences in the Braaten-Pisarski-Kapusta (BPK) series
in powers of $g^{m+2n} \ln^n g$, but it smoothly incorporates
the case of nonzero chemical potential(s).
The latter is a powerful nonperturbative tool to calculate
equations of state for both phases. However, up to now, there
are no realistic lattice data available for nonzero chemical
potential(s) (for problems to introduce it on the
lattice see, for example, recent paper [9]).
 As it was emphasized in Ref. [10], at present the phase
transition at finite baryon chemical potential can be only studied
within the phenomenological models. The most popular among them
are, of course, the bag-type models [11], which differ from each
other by modelling the equation of state of the hadronic
phase [12]. The
equation of state of the QGP phase is usually approximated
by the thermal perturbation theory as an ideal (noninteracting)
gas consisting of gluons and massless quarks. The
generalizations on massive quarks and interacting QGP are also
possible though this is not a simple task because of the
nonperturbative contributions to the thermodynamic potential,
coming from the scales $gT$ and $g^2T$,
which badly affect the above mentioned BPK series.
 
  Within the framework of the bag-type models, a phase transition
between the hadronic gas (HG) and the QGP phases is constructed
via the Gibbs criteria for a phase equilibrium. So the phase
transition is necessarily of the first order. This means that the
thermodynamic quantities of interest are discontinuous
across the critical curve. The entropy content of the QGP is much
higher than in the hadronic phase, so the transition from QGP to
HG phase is impossible at fixed temperature and chemical
potential(s), since in physical processes like the heavy ion
collisions the entropy cannot decrease. To make
the phase transition reversible, it is necessary thus to change
the
values of temperature and chemical potential(s) in order to ensue
the Gibbs
conditions for the phase transition. In the above mentioned paper
[10] another interesting way was proposed to construct a
continuous
transition which preserves the Gibbs criteria but leads to a
temperature and chemical potential dependent bag constant. In
particular, it is explicitly shown that the specific entropy per
baryon, which is an important physical observable since
it indicates the
strangeness production in heavy ion collisions, becomes, in this
case, continuous across the phase boundary.
 
  We have recently proposed a new bag-type model, the
so-called extended bag model [13]. It complements the standard
bag model by an Ansatz for the pressure at the phase
boundary (measured
in terms of the bag constant) which allows one to correctly
take into account the nonperturbative vacuum effects from both
sides of the equilibrium condition. Our main goal in this paper is
to show that our model automatically provides an isentropic
equilibrium deconfining phase transition with a
temperature and chemical potential dependent bag pressure at the
phase boundary. Thus our model makes it feasible to
regulate the total entropy and baryon number in both phases, i. e.
the flow of the specific entropy per baryon across the phase boundary.
 
  II. For the reader's convinient and for the sake of
future application, let us briefly discuss the main features
of our model in the case of the noninteracting QGP.
The QGP state equation determines the dependence of the QGP
thermodynamical quantities such as the energy density $\epsilon$
and pressure $P$
on the thermodynamical variables temperature $T$ and quark
chemical potentials $\mu_f$. There exist excellent reviews on the physics
of the QGP (see, for example, Refs. [14-16]), as well as on the
phase transitions in it [17]. The QGP pressure
(i. e., the thermodynamic potential $\Omega$, apart from the
sign) is [16]
\begin{equation}
P = {1 \over 3} f_{SB} T^4 + {N_f \over 2}
\mu_f^2 T^2 +{N_f \over 4 \pi^2} \mu_f^4 - B,
\end{equation}
where $B$ is the bag constant (see below), while $N_f$ is the number
of different quark flavours. In what follows we will consider the
values $N_f = 0, 1, 2$ since the inclusion of the strange $(s)$
quark requires a special treatment [18].
The value $N_f = 0$ describes the case of the pure gluon plasma.
Note also that the state equation (1) is derived by
neglecting quark current masses.
 
  The constant $f_{SB}$, entering the equation
of state (1),
\begin{equation}
f_{SB} \equiv f_{SB} (N_f) = f_0 (N_f) =
{\pi^2 \over 5} \Bigl({8 \over 3} + {7 \over 4} N_f \Bigr)
\end{equation}
is the Stefan-Boltzmann (SB) constant, which determines
the ideal (noninteracting gluons and massless quarks) gas limit.
Obviously, for $N_f=0$ it equals to the standard SB
constant of the ideal gluon gas.
Note that the temperature
dependence of the QGP equation of state is dominant over the
dependence on quark chemical potentials.
 
 The energy density $\epsilon$ of the noninteracting
QGP can be obtained from the thermodynamic potential (1)
as follows ($P = - \Omega$)
\begin{equation}
\epsilon = 3 P + 4 B,
\end{equation}
so the bag constant determines, in general, deviation from
the ideal gas relation between pressure and energy density.
Let us make a few remarks in advance. Our calculations
for the noninteracting QGP are not based on the bag model state
equation (3) [11, 17]. The constraint, determining
the phase transition, will be obtained with the help of the Ansatz
which is beyond the bag model and it is general (see below,
part III and Ref. [13]). We will use
numerical values for the bag constant which were obtained from a
completely different source, namely they were calculated in
the zero modes enhancement (ZME) model of the true QCD vacuum
taking the instanton contributions into account as well [2, 13].
 
III. The Gibbs conditions for the phase equilibrium between
hadronic gas (HG) and QGP phases at $T=T_c$ are formulated as
follows [11]
\begin{equation}
P_h = P_q = P_c; \quad T_h = T_q = T_c; \quad
3 \mu_f = \mu_c,
\end{equation}
where subscripts h, q and c refer to HG, QGP phases and
transition (critical or crossover) region, respectively. At the
same time, the difference between $\epsilon_q - \epsilon_h$ at
$T=T_c$ can remain finite (nonzero) and determines
the latent heat (LH), $\epsilon_{LH}$. Let us remind that $P_q$
and $\epsilon_q$ are determined by Eqs. (1) and (3), respectively.
 
  Let us formulate our primary assumption (Ansatz) now.
The state equation for
the hadronic phase (the left hand side of the equilibrium
condition (4)) is strongly model dependent [11, 12]. However,
in any model the pressure $P_h$ at any values of temperature $T$
and baryonic chemical potential $\mu$,
in particular at $T=T_c$ and $\mu = \mu_c$ , can be measured
in terms of the above mentioned bag constant, i.e. let us put
\begin{equation}
 P_h (T_c, \mu_c) = { b_h \over a_h + N_f} B,
\end{equation}
where the so-called parametric functions $b_h \equiv b_h (T_c,
\mu_c)$
and $a_h \equiv a_h (T_c, \mu_c)$ describe the details of the HG
phase at the phase boundary. Evidently, they may only depend
on the set of independent dimensionless variables
which characterize the HG phase. For example,
\begin{eqnarray}
a_h \equiv a_h (T_c, \mu_c) = a_h (x_c, t_c, y_c, z_c), \nonumber\\
b_h \equiv b_h (T_c, \mu_c) = b_h (x_c, t_c, y_c, z_c),
\end{eqnarray}
where
\begin{equation}
x_c = {\mu_c \over T_c}, \quad  t_c = {\tilde{B}^{1/4} \over T_c},
\quad y_c = {\mu_c \over m}, \quad z_c = \mu_c R_0,
\end{equation}
and $m$ denotes the hadron mass while $R_0$ denotes the radius of
the nucleon, so it allows one to take into account finite size
effects due to hard core repulsion between nucleons (extended
volume corrections) [19]. These variables are independent and all
other possible dimensionless variables are obtained by combination
of these, for example, $R_0 T_c = z_c / x_c, \quad m / T_c =x_c /
y_c, \quad \mu_c / \tilde{B}^{1/4} = x_c / t_c, etc$. Also the set
of independent dimensionless variables at the phase boundary (7)
may be extended in order to treat the HG phase in a more
sophisticated way. However, in any case, the parametric
functions should be symmetric, i. e.
$a_h (T_c, \mu_c)= a_h (- T_c, - \mu_c)$ and
$b_h (T_c, \mu_c)= b_h (- T_c, - \mu_c)$.
 
From
Eqs. (1) and (4-5) at $T=T_c$ and $\mu_f = \mu_c / 3$, one obtains
\begin{equation}
f_{SB}(N_f) T^4_c + {N_f \over 6} \mu_c^2 T^2_c
+ {N_f \over 108 \pi^2} \mu_c^4 = {3 \over a_h + N_f} \tilde{B},
\end{equation}
where we introduced a new "physical" (effective) bag constant
as follows
\begin{equation}
\tilde{B} =  (b_h + a_h + N_f) B,
\end{equation}
and it linearly depends on $N_f$, as it should be, at log-loop
level (see our papers [2] and Ref. [13]).
In connection with our Ansatz (5) a few remarks are in order.
The alternative parametrization with respect to $N_f$, namely
$P_h (T_c, \mu_c) = (b_h^{'} / a_h^{'} N_f + 1)B$ leads to the
effective bag constant as $\tilde{B} =(b_h^{'} + a_h^{'}N_f +1)B$.
The parameteric functions $b_h$ and $a_h$, as well as $b_h^{'}$
and $a_h^{'}$, as was mentioned above, may, in
principle, arbitrarily depend on dimensionless variables (7).
If, for example, $a_h^{'}$ vanishes at
$\mu_c =0$ then the linear dependence of $\tilde{B}$ on $N_f$ will
be spoiled. In other words, the choosen parametrization guarantees
the linear dependence of $\tilde{B}$ on $N_f$ and the
alternative one does not.
 
From now on,  $T_c$
and $\mu_c$ will be calculated in terms of $\tilde{B}$ and not
that of old $B$, i. e. a definite numerical value
will be assigned to $\tilde{B}$.
So we consider $\tilde{B}$ as the physical bag constant,
while $B$ as an unphysical "bare" one.
The bag constant is universal and it represents the complex
nonperturbative structure of the QCD true vacuum. Thus the
proposed Ansatz, allows one to treat nonperturbative
vacuum effects (which are parametrized in terms of $\tilde{B}$)
from both sides of the equilibrium condition (4).
However, it still remains dependent on the arbitrary
parameter $a_h$. In order to eliminate this dependence, let us
normalize the thermodynamic potential at the phase boundary in Eq.
(8) to the standard SB constant (2). This yields
\begin{equation}
\tilde{f}_{SB} (N_f) = (N_f + a_h) f_{SB} (N_f) = f_{SB}(0) \quad
at \quad N_f = 0,
\end{equation}
and one immediately arrives at $a_h \equiv a_h (T_c, \mu_c) = 1$.
This is our normalization condition and it leads to good numerical
results for $T_c$ and $\mu_c$ (see Ref. [13]).
So the general constraint (8) finally becomes
uniquely determined, namely (Fig. 1)
\begin{equation}
f_{SB} T^4_c + {N_f \over 6} \mu_c^2 T^2_c
+ {N_f \over 108 \pi^2} \mu_c^4 = {3 \over N_f + 1} \tilde{B},
\end{equation}
and consequently allows one to investigate the bulk
thermodynamic quantites in the vicinity of a critical point.
Let us emphasize the important observation that the numerical
values of $T_c$ and $\mu_c$ calculated from the constraint (11)
do not depend on the way how one approximates the equation of
state
of the hadronic phase. In contrast to the standard bag-type models
(differed from each other by modelling the hadronic phase [12]),
in the
extended bag model their values depend only on $\tilde{B}$ which
incorporates nonperturbative vacuum effects from both sides of the
Gibbs equilibrium condition (4) as it has been already emphasized
above.
 
 The QGP energy density and pressure, however, remain undetermined
with our Ansatz (5) at this stage. In terms of $\tilde{B}$ they become
\begin{equation}
P_h (T_c) = P_q (T_c) ={b_h \over (N_f + 1)[(N_f + 1)+ b_h ]}
\tilde{B}
\end{equation}
and because of Eq. (3),
\begin{equation}
\epsilon_q (T_c) \equiv \epsilon_c = {3 b_h + 4 (N_f + 1) \over
(N_f +1) [(N_f +1) + b_h]} \tilde{B}.
\end{equation}
As mentioned above, the
unknown $b_h$ reflects the fact that the hadronic phase
state equation is strongly model dependent. The unknown $b_h$
is the price one pays to determine the above mentioned physical
quantities with the help of our Ansatz. Precisely for this reason,
the bag model state equation (3) plays no role in our numerical
investigation of the phase transition with the constraint (11)
from which $T_c$, as well as $\mu_c$, can be derived.
Concluding this part, let us note
that our model cetainly requires the coexistence regime between
QGP and HG phases since $\epsilon_q (T_c)$, as given by Eq. (13),
explicitly
depends on $b_h$ which describes details of the HG phase at
$T=T_c$.
 
  IV. Let us discuss now some general features of the extended bag
model. The bare bag constant $B$ in Eq. (9) as well as the
hadronic gas pressure $P_h$ in Eq. (12) were determined at the
phase boundary only, namely
\begin{equation}
B (T_c, \mu_c) = {\tilde{B} \over b_h (T_c, \mu_c) + a_h (T_c,
\mu_c) + N_f }
\end{equation}
and
\begin{equation}
P_h (T_c, \mu_c) = {b_h (T_c, \mu_c) \over (a_h (T_c, \mu_c) +
N_f)} { \tilde{B} \over [ b_h (T_c, \mu_c) + a_h (T_c, \mu_c) +
N_f]},
\end{equation}
respectively. In both equations the normalization condition
$a_h (T_c, \mu_c) = 1$ is to be used.
Note that in these expressions (in comparison with Eqs. (9) and
(12)) we have only restored the dependence on $T_c$ and $\mu_c$,
i. e. using relations (6). Let us emphasize that
$b_h$ in general is not constant. Only for a simple hadronic phase
(consisting, for example, of only massless pion gas) it is
constant [13].
The dependence of the QGP pressure $P_q$ (1) on $T$ and $\mu$
remains, obviously, in this case unchanged, since the
derivatives of bare $B$ (14) with respect to $T$ and $\mu$
disappear. The phase transition with the bag pressure determined
by Eq. (14) was investigated in our previous work [13].
 
One may go outside the phase boundary in our model as well by
considering a $T$ and $\mu$
dependent parametric functions, $b_h (T, \mu)$ and $a_h (T, \mu)$,
which, at the phase boundary, become $b_h$ and 1. So extrapolation
of the bag pressure (14) outside the phase boundary
is given by
\begin{equation}
B (T, \mu) = {\tilde{B} \over b_h (T, \mu) + a_h (T, \mu)+ N_f }.
\end{equation}
Hence the HG pressure (15) also becomes dependent on
$T$ and $\mu$ determining it outside the phase boundary as
well, i. e.
\begin{equation}
P_h (T, \mu) = {b_h (T, \mu) \over (a_h (T, \mu) + N_f)}
{ \tilde{B} \over [b_h (T, \mu) +a_h (T, \mu) + N_f]},
\end{equation}
which, evidently, at the phase boundary becomes Eq. (12) or Eq.
(15).  The QGP pressure $P_q$ (1) becomes
\begin{equation}
P_q = {1 \over 3} f_{SB}(N_f) T^4 + {N_f \over 2} \mu_f^2 T^2
+ {N_f \over 4 \pi^2} \mu_f^4 -
{\tilde{B} \over b_h (T, \mu) + a_h (T, \mu) + N_f }.
\end{equation}
 A simple relation (3) between the energy density and pressure now
does
not take place. The QGP energy density should be deteremined from
a general thermodynamic relation, namely
\begin{equation}
\epsilon_q =  T \Bigl( {\partial P_q \over \partial T}
\Bigr)_{\mu_f}  + \mu_f \Bigl( {\partial P_q \over \partial \mu_f}
\Bigr)_T - P_q,
\end{equation}
where $P_q$ is given by the previous equation. Then one obtains
\begin{eqnarray}
\epsilon_q = 3 P_q &+& 4 {\tilde{B} \over b_h (T, \mu) + a_h (T,
\mu) + N_f}  \nonumber\\
&+& {\tilde{B} \over (b_h (T, \mu) + a_h (T, \mu) + N_f
)^2} \Bigl( T {\partial \over \partial T}
+ \mu {\partial \over \partial \mu} \Bigr) \Bigl( b_h (T, \mu)
+ a_h (T, \mu) \Bigr),
\end{eqnarray}
where we replaced $\mu_f$ by $\mu /3$.
From this equation one recovers the standard relation (3) when the
bag constant $B$ does not depend on $T$ and $\mu$.
In the same way should be determined the hadronic energy density
\begin{equation}
\epsilon_h =  T \Bigl( {\partial P_h \over \partial T}
\Bigr)_{\mu}  + \mu \Bigl( {\partial P_h \over \partial \mu}
\Bigr)_T - P_h,
\end{equation}
where $P_h$ is now given by Eq. (17). One finally obtains
\begin{eqnarray}
\epsilon_h  &=&
{ \tilde{B} \over (b_h (T, \mu) + a_h (T, \mu) + N_f)^2}
\Bigl( T {\partial \over \partial T} + \mu { \partial
\over \partial \mu} \Bigr) b_h (T, \mu) \nonumber\\
&-& { \tilde{B} b_h (T, \mu) [ b_h (T, \mu) + 2 a_h (T, \mu) + 2
N_f] \over (a_h (T, \mu)+ N_f)^2
(b_h (T, \mu) + a_h (T, \mu) + N_f)^2}
\Bigl( T {\partial \over \partial T}
+ \mu {\partial \over \partial \mu} \Bigr) a_h (T, \mu) - P_h.
\end{eqnarray}
Note, Eq. (22) need not be multiplied by the overall factor which
takes into account extended volume corrections [13, 19] since our
parametric functions automatically incorporate them (see Ref. [13]
and below).
Combining this equation with Eq. (20) at the phase
boundary, for the latent heat
$\epsilon_{LH} = \epsilon_c  - \epsilon_h$ one obtains
\begin{equation}
\epsilon_{LH} = {4 \over N_f+1} \tilde{B}
+ { \tilde{B} \over ( N_f +1)^2 }
\Bigl( T {\partial \over \partial T} + \mu { \partial
\over \partial \mu} \Bigr) a_h (T, \mu),
\end{equation}
where, after taking the derivatives, one needs to put $T=T_c$ and
$\mu=\mu_c$. In derivation of Eq. (23), we have used
$P_h = P_q$ and $\epsilon_q \equiv \epsilon_c$ at the phase boundary
$T=T_c$ and $\mu= \mu_c$ and also the explicit expression (17)
which becomes Eq. (15). Thus to leading order the latent heat, as
given by Eq. (23), does not
depend on the details of the hadronic phase which are described by
the parametric functions $b_h (T, \mu)$ and $a_h (T, \mu)$.
Note that the second (next-to-leading) term is suppressed
approximately by one order of magnitude in comparison with the first
(leading) term (especially for the physically relevant case of the
two light quark species $N_f=2$). The parametric function
$a_h(T, \mu)$ is a slowly varying  function (it is nearly
constant in the vicinity of the phase transition due to its above
mentioned normalization condition).
 
Using now our values of the effective bag constant $\tilde{B}$
[13], one obtains the numerical values of the latent heat, up to
leading order, in our model as follows
\begin{eqnarray}
\epsilon_{LH} (N_f = 0) &=& 3.12 \ GeV / fm^3, \nonumber\\
\epsilon_{LH} (N_f = 1) &=& 1.82 \ GeV / fm^3, \nonumber\\
\epsilon_{LH} (N_f = 2) &=& 1.38 \ GeV / fm^3.
\end{eqnarray}
 At the same time, the QGP energy density $\epsilon_q$ (20)
depends explicitly on the parametric functions $b_h (T, \mu)$ and
$a_h (T, \mu)$. Since they
describe the details of the hadronic phase, so this requires
the coexistense regime between QGP and HG phases in our model as
it was already mentioned above.
 
 Concluding this part of our work, let us remind the reader that
the
constraint condition (11), determining the critical curve, does
not depend on how one goes outside the phase boundary. Thus
the critical temperature $T_c$ and the critical chemical potential
$\mu_c$ do not depend on the details of the hadronic phase which
are decribed by the parametric functions $b_h (T, \mu)$ and
$a_h (T, \mu)$. In our model their numerical values [13]
are mainly determined by the vacuum effects in both phases.
 
 V. It was proposed in Ref. [10] to construct the isentropic
equilibrium phase transition from QGP to HG by considering a $T$
and $\mu$ dependent bag pressure. Due to the above general
discussion, this can be easily incorporated in our model.
The isentropic equilibrium phase transition from QGP to a hadronic
gas at fixed $T$ and $\mu$ via the ratio specific entropy per
baryon is determined as
\begin{equation}
\Bigl({s \over n_B } \Bigr)_q = \Bigl({s \over n_B } \Bigr)_h,
\end{equation}
where $s$ and $n_B$ are the entropy and baryon number densities,
respectively. For the QGP phase they are defined as
\begin{equation}
s_q = \Bigl( {\partial P_q \over \partial T} \Bigr)_{\mu_f}
\end{equation}
and
\begin{equation}
n_B = {1 \over 3} \Bigl( {\partial P_q \over \partial \mu_f}
\Bigr)_T.
\end{equation}
 
  In the HG phase these quantities are defined as follows
\begin{equation}
s_h = \Bigl( {\partial P_h \over \partial T} \Bigr)_{\mu}.
\end{equation}
and
\begin{equation}
n_B = \Bigl( {\partial P_h \over \partial \mu} \Bigr)_T.
\end{equation}
 
 In the extended bag model the parametric functions $b_h (T, \mu)$
and $a_h (T, \mu)$ can only be the functions of the dimensionless
variables, i. e
\begin{eqnarray}
a_h (T, \mu) &=& a_h (x, t, \tau, y, z, q),
\nonumber\\
b_h (T, \mu) &=& b_h (x, t, \tau, y, z, q),
\end{eqnarray}
where
\begin{equation}
x = {\mu \over T}, \quad t = { \tilde{B}^{1/4} \over T},
\quad  \tau = {T \over T_c}, \quad y = {\mu \over m}, \quad z =
\mu R_0, \quad q = {\mu \over \mu_c}.
\end{equation}
Evidently, this extrapolates the above introduced set of
independent dimensionless variables (6-7) outside the phase
boundary. These six variables are independent and all other
possible dimensionless variables are obtained by a combination of
these, for example, $R_0 T = z/x$, ${m \over T} = x/y$,
${\mu \over \tilde{B}^{1/4}} = x/t$, etc. Note that the solutions
should be symmetric, i. e. $b_h (T, \mu) = b_h (-T, -\mu)$ and
$a_h (T, \mu) = a_h (-T, -\mu)$ because of the corresponding
symmetry in the pressure. Also, the set of
independent dimensionless variables (31) may be extended, treating
the hadronic phase in a more sophisticated way, etc. However, for
our main purpose in this work (isentropic equilibrium phase
transition) this set of variables is completely sufficient.
 
 The HG and QGP pressures are given by
 Eqs. (17) and (18), respectively. It is convenient to
introduce the functions defined as
\begin{equation}
f \equiv f (T, \mu) =  b_h (T, \mu) + a_h (T, \mu) + N_f = f (x,
t, \tau, y, z, q).
\end{equation}
and
\begin{equation}
g \equiv g (T, \mu) =  {b_h (T, \mu) \over a_h (T, \mu) + N_f} =
g (x, t, \tau, y, z, q).
\end{equation}
Then from definitions (26) and (27), one obtains
\begin{equation}
s_q = {4 \over 3}f_{SB}(N_f) T^3 + {N_f \over 9} \mu^2 T -
{ \partial \over \partial T} {\tilde{B} \over f (T, \mu) },
\end{equation}
and
\begin{equation}
n_B = {N_f \over 9} \mu T^2
+ {N_f \over 81 \pi^2} \mu^3 - { \partial \over \partial \mu}
{\tilde{B} \over f(T, \mu) },
\end{equation}
respectively. Here we again replaced $\mu_f$ by $\mu /3$.
 
The entropy and baryon number conservation condition (25), on
account of relations (34-35) and Eq. (17) on account of
definitions (32-33), finally becomes
\begin{equation}
{ T^3 s_0 f^2 + \tilde{B} {\partial f \over \partial T} \over
T^3 n_0 f^2 + \tilde{B} {\partial f \over \partial \mu}} =
{ f {\partial g \over \partial T} - g {\partial f \over \partial T}
\over f {\partial g \over \partial \mu} - g {\partial f \over
\partial \mu}},
\end{equation}
where
\begin{eqnarray}
s_0 &=& {4 \over 3}f_{SB}(N_f) + {N_f \over 9} (\mu / T)^2, \nonumber\\
n_0 &=& {N_f \over 9} (\mu / T)
+ {N_f \over 81 \pi^2} (\mu / T)^3.
\end{eqnarray}
This
relation (36) itself is completely sufficient to guarantee
the isentropic equilibrium transition (since it is only one
relation for two unknown functions) but it is not sufficient
to determine both functions $f$ and $g$ and hence the functional
dependence of $b_h$ and $a_h$ on $T$ and $\mu$. One needs a second
independent relation between them. We suggest the
QGP fireball condition $ (s/n_B)_q \simeq 50$ [20] as such.
In terms of functions $f$ and $g$ this is
\begin{equation}
{ T^3 s_0 f^2 + \tilde{B} {\partial f \over \partial T} \over
T^3 n_0 f^2 + \tilde{B} {\partial f \over \partial \mu}} = 50.
\end{equation}
Thus one obtains a system of two strongly
coupled nonlinear differential equations
in partial derivatives for $f$ and $g$ parametric functions
which, of course, requires a separate consideration.
 
  VI. It is instructive, however, to investigate a simplified
case which allows us to compare our results with those obtained
in Refs. [10] and [21]. As mentioned above, in Ref. [10] the
isentropic equilibrium condition was constructed by extrapolating
outside the phase boundary the bag pressure (16) only.
 In terms of our parametric functions $f$ and $g$ (32-33), this
means that the dependence on the parametric function $g$ in the
isentropic equilibrium condition (36) is neglected. Then
finally the condition (36) becomes
\begin{equation}
{ s_0 \over n_0 } =
{ {\partial \over \partial T} f \over
{\partial \over \partial \mu} f}.
\end{equation}
In terms
of the dimensionless variables (31), from Eq. (39) one obtains
\begin{equation}
x n_0 \tau {\partial f \over \partial \tau}
-x n_0 t {\partial f \over \partial t} - x (s_0 + x n_0)
{\partial f \over \partial x} =
s_0 \Bigl( y {\partial f \over \partial y}
+ z {\partial f \over \partial z} +q {\partial f \over
\partial q} \Bigr),
\end{equation}
where $s_0$ and $n_0$ are functions of variable $x$ only and the
explicit expressions can be easily obtained from relations (37) on
account of the above mentioned substitution (31). It is easy to
check that the general solution of this equation is
\begin{equation}
f (x, t, \tau, y, z, q) = A x^{-C} t^{C_2}
\tau^{C_2-C_1} y^{C_3} z^{C_4} q^{C-C_3-C_4}
\Bigl( 1 + a_1 x^2 + a_2 x^4 \Bigr)^{{C-C_1 \over 4}},
\end{equation}
where
\begin{equation}
a_1 = {N_f \over 6 f_{SB}(N_f)},
\quad a_2 = { N_f \over 108 \pi^2 f_{SB}(N_f)}.
\end{equation}
Below, for simplicity, all dimensional as well as
dimensionless constants, appearing
in the solution (41), will be always included in the integration
"constant" $A$ and the same notation will be retained. Constants
$C's$ are
the constants of separation. In terms of $T$ and $\mu$ it becomes
\begin{equation}
f (T, \mu) = b_h (T, \mu) + a_h (T, \mu) + N_f = A
\mu_c^{C_3+C_4-C} T_c^{C_1-C_2} T^{C-C_1}
\Bigl( 1 + a_1 x^2 + a_2 x^4 \Bigr)^{{C-C_1 \over 4}}.
\end{equation}
 At the phase boundary $x=x_c$, on account of the constraint (11),
it finally becomes
\begin{equation}
f (T_c, \mu_c) = b_h (T_c, \mu_c) + N_f + 1  = A \mu_c^{C_3+C_4-C}
T_c^{C_1-C_2},
\end{equation}
where we used the normalization condition $a_h (T_c, \mu_c) = 1$.
This expression should be finite at the end points $(T_c, 0)$
and $(0, \mu_c)$ of the phase diagram $(T_c, \mu_c)$ shown in Fig.
1, in order to make the hadronic pressure at the phase boundary
also finite as it should be. Thus the solution (43) becomes
\begin{equation}
f (T, \mu) = b_h (T, \mu) + a_h (T, \mu) + 1 = A T^{C-C_1}
\Bigl( 1 + a_1 x^2 + a_2 x^4 \Bigr)^{{C-C_1 \over 4}}.
\end{equation}
So the bag pressure (16) at the phase boundary becomes
simply $ B = \tilde{B} / f(T_c, \mu_c)$, i. e. it remains
finite but arbitrary (since the "constant" $A$ may depend on
$T_c$ and $\mu_c$). Nevertheless the
condition (25) remains continuous across the phase boundary since
the ratio in the right hand side of the isentropic relation (39)
remains, evidently, finite in the limit $x \rightarrow x_c = \mu_c
/ T_c$. This is
the result of that feature of our model that
nonperturbative vacuum effects from both
sides of the equilibrium condition are taken care of.
Apparently, this makes
it possible to regulate the total entropy and baryon number in
both phases, i.e. the flow of the specific entropy per baryon
across the phase boundary.
 
 VII. It is easy to see that the asymptotics of the solution (45)
outside the phase boundary are determined by the separation constant
$C-C_1$ and they do not dependend on the variables $y$, $z$ and
$q$. Indeed, at $x \rightarrow 0$, which means hot HG, at
fixed
$T$) and low baryon density ($\mu \rightarrow 0$), from (45) one
obtains that the value of $\tilde{B}$ in this limit is determined
by fixed temperature and $C-C_1$, i. e. becomes constant, $A_0$.
In the same way at $x \rightarrow \infty$, which means
cold ($ T\rightarrow 0$), baryon dense matter (at fixed
$\mu$), from (45) one obtains that its value in this limit now
is determined by fixed $\mu$ and again $C-C_1$, i. e. tends to
constant, $A_{\infty}$.
Then the qualitative behaviour of the bag constant $B$ (16) as a
function of $x$ (or $T$ and $\mu$) is shown in Fig. 2.
 
  Thus in our model the bag pressure always remains finite and
positive (for the simplified case this is certainly so and
apparently this will be valid for the general case as well) in
accordance with its definition as the difference between the
perturbative (normalizable to zero) and the nonperturbative vacua.
This definition
remains valid also for a $T$ and $\mu$ dependent bag constant.
In the
relevant ranges of $T$ and $\mu$, the bag pressure, as obtained in
Ref. [10], is slowly varying function, i. e. it is nearly
constant.
This can be seen from Figs. 1 and 2 presented in Ref. [21].
This behaviour of the bag pressure is in agreement with our
model as shown in Fig. 2. In the above mentioned paper [21], this
behaviour [10] was criticized since it increases with increasing
chemical potential at a fixed $T$ and thus defies a physical
interpretation. However, the finite zeros of the bag constant
(especially as a function of $T$) obtained
and shown in Figs. 1 and 2 of the above mentioned paper
[21] are completely unsatisfactory since they strictly contradict
the definition of the bag constant. As a function
of $T$ and $\mu$, it can only tend to zero (in the worst case)
but never assume zero, i. e. it must always remain positive.
These unphysical finite zero points are artifacts of the
inconsistent
approximation used in that paper. Firstly, it is incorrect
to directly compare results obtained at different scales of the
QCD coupling constant, $\alpha_s =0$ and the running $\alpha_s$.
The former should be treated at an infinitely large scale while
the
latter one - at finite, fixed scale, so there is no smooth limit
to $\alpha_s =0$. That is why the noninteracting QGP should be
investigated separately from the interacting one. Secondly, and
more
importantly, it is well known that the first
nonperturbative
contribution of order $\alpha_s^{3/2}$ to the thermodynamical
potential is uncomfortably large even for small $\alpha_s$ [22].
So the result obtained on account of only the first perturbative
correction of order $\alpha_s$ will be completely destorded when
the above mentioned nonperturbative contribution will be
incorporated in the
calculation of the thermodynamic potential.
Thus to treat the noninteracting QGP from the very beginning
is a selfconsistent approximation [10, 13] while it is not like
that to
treat the interacting QGP up to $\alpha_s$ order only [21].
In order to achieve that the approximation should be
selfconsistent for a running $\alpha_s$ as well, it is
necessary to consider at least the nonperturbative
contributions from the scale $gT$. Finally, in Ref. [21], in
comparison with Ref. [10] and present investigation, there is
no phase boundary at all across which entropy per baryon may be
continuous on account of a $T$- and $\mu$- dependent bag constant.
 
The extension of our model to the case of a running coupling
constant (which has been already very tentatively treated in
Ref. [23]) and the determination of the nontrivial fluctuations in
the interacting QGP will be the subject of the subsequent paper.
 
  The authors would like to thank J. Zim\'anyi,
K. T\'oth, Gy. Kluge, T. Bir\'o, P. L\'evai, and T. Cs\"org\H o
for useful discussions, remarks and support.

\vfill
\eject

\vfill
\eject

\begin{figure}
 
\caption{The phase diagram in the plane $(T_c, \mu_c)$ measured
         in inits of $MeV$. Here the
         curve is shown for the physically relevant case of
         the two light quarks $N_f=2$ and the value
$\tilde{B}^{1/4} = 300 \ MeV$ was used (see Ref. [13]). }
 
\bigskip
 
\caption{The qualitative bahaviour of the bag pressure (16) as a
function of $x= \mu / T$. The upper line corresponds to
$A_0 > A_{\infty}$, while the lower one to - $A_0 < A_{\infty}$.}
\end{figure}


\begin{references}
\bibitem{1}
   H.H.Gutbrod and J.Rafelski, in: "Particle Production
   in Highly Excited Matter", (Ed. by H.H.Gutbrod and J.Rafelski,
   NATO ASI Series B: Physics Vol.303), p.1
\bibitem{2}
   V.Gogohia, Gy.Kluge and M. Priszny\'ak, Phys.Lett., {\bf B368}
   (1996) 221; Phys.Lett., {\bf B378} (1996) 385;
   hep-ph/9509427
\bibitem{3}
   A.Chodos et al., Phys.Rev., {\bf D9} (1974) 3471
\bibitem{4}
   T.DeGrand, R.L.Jaffe, K.Johnson and J.Kiskis,
   Phys.Rev., {\bf D12} (1975) 2060
\bibitem{5}
   E.Braaten and A.Nieto, Phys.Rev., {\bf D53} (1996) 3421; \\
   E.Braaten and A.Nieto, Phys.Rev.Lett., {\bf 76} (1996) 1417
\bibitem{6}
   E.Braaten and R.D.Pisarski, Nucl.Phys., {\bf B337} (1990) 569
\bibitem{7}
   J.I.Kapusta, Finite Temperature Field Theory (Cambr.
   Univ. Press, England, 1989)
\bibitem{8}
   H.J.Rothe, Lattice Gauge Theories, An Introduction (WS,
   Lect. Notes in Phys. - Vol. 43, 1992); I.Montvay and
   G.M\"unster, Quantum Fields on a Lattice (Cambr., NY, 1994)
\bibitem{9}
   M.-P.Lombardo, J.B.Kogut and D.K.Sinclair, Phys.Rev.,
   {\bf D54} (1996) 2203
\bibitem{10}
   A.Leonidov, K.Redlich, H.Satz, E.Suhonen and G.Weber,
   Phys.Rev., {\bf D50} (1994) 4657
\bibitem{11}
   J.Cleymans, R.V.Gavai and E.Suhonen, Phys.Rep., {\bf 130}
   (1986) 217
\bibitem{12}
   L.Csernai, Introduction to Relativistic Heavy Ion
   Collisions, (J. Wiley and Sons, 1994);
   K.S.Lee, M.J.Rhoades-Brown and U.Heinz, Phys.Rev., {\bf D37}
   (1988) 1452
\bibitem{13}
   V.Gogohia, B.Luk\'acs and M. Priszny\'ak, hep-ph/9604308
\bibitem{14}
   H.Satz, Preprint CERN-TH.7410/94, BI-TP 94/45
\bibitem{15}
   L.McLerran, Rev.Mod.Phys., {\bf 58} (1986) 1021
\bibitem{16}
   B.M\"uller, Rep.Prog.Phys., {\bf 58} (1995) 611
\bibitem{17}
   H.Meyer-Ortmanns, Rev.Mod.Phys., {\bf 68} (1996) 473
\bibitem{18}
   J.Zim\'anyi, P.L\'evai, B.Luk\'acs and A.R\'acz, in: "Particle
   Production
   in Highly Excited Matter", (Ed. by H.H.Gutbrod and J.Rafelski,
   NATO ASI Series B: Physics Vol.303)
\bibitem{19}
   J.Cleymans, K.Redlich, H.Satz, and E.Suhonen,
   Z.Phys. C- Part. and Fields, {\bf 33} (1986) 151
\bibitem{20}
   J.Rafelski, in: "Particle Production
   in Highly Excited Matter", (Ed. by H.H.Gutbrod and J.Rafelski,
   NATO ASI Series B: Physics Vol.303), p.529
\bibitem{21}
   B.K.Patra and C.P.Singh, Phys.Rev., {\bf D54} (1996) 3551
\bibitem{22}
   E.V.Shuryak, Phys.Rep., {\bf 115} (1984) 151
\bibitem{23}
   V.Gogohia, B.Luk\'acs and M. Priszny\'ak, Preprint KFKI-1996-14/A
\end{references}
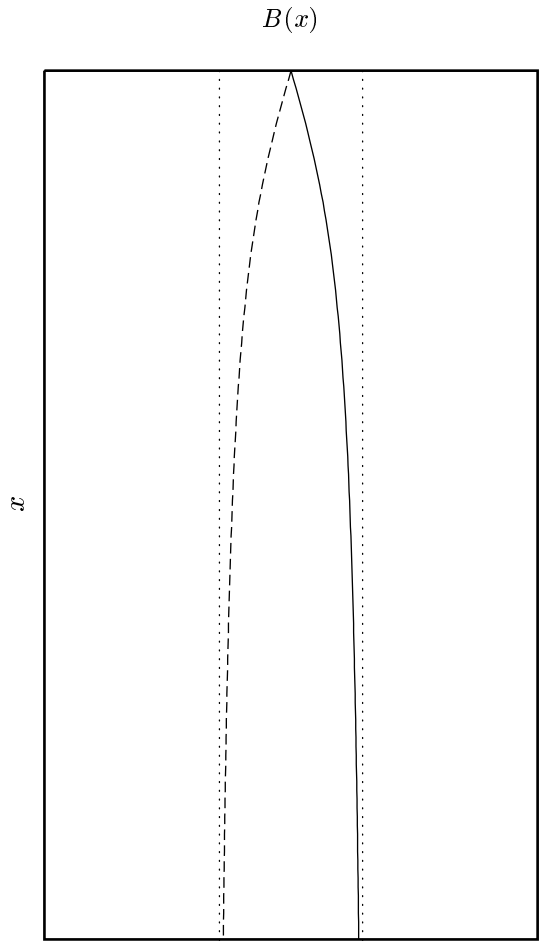
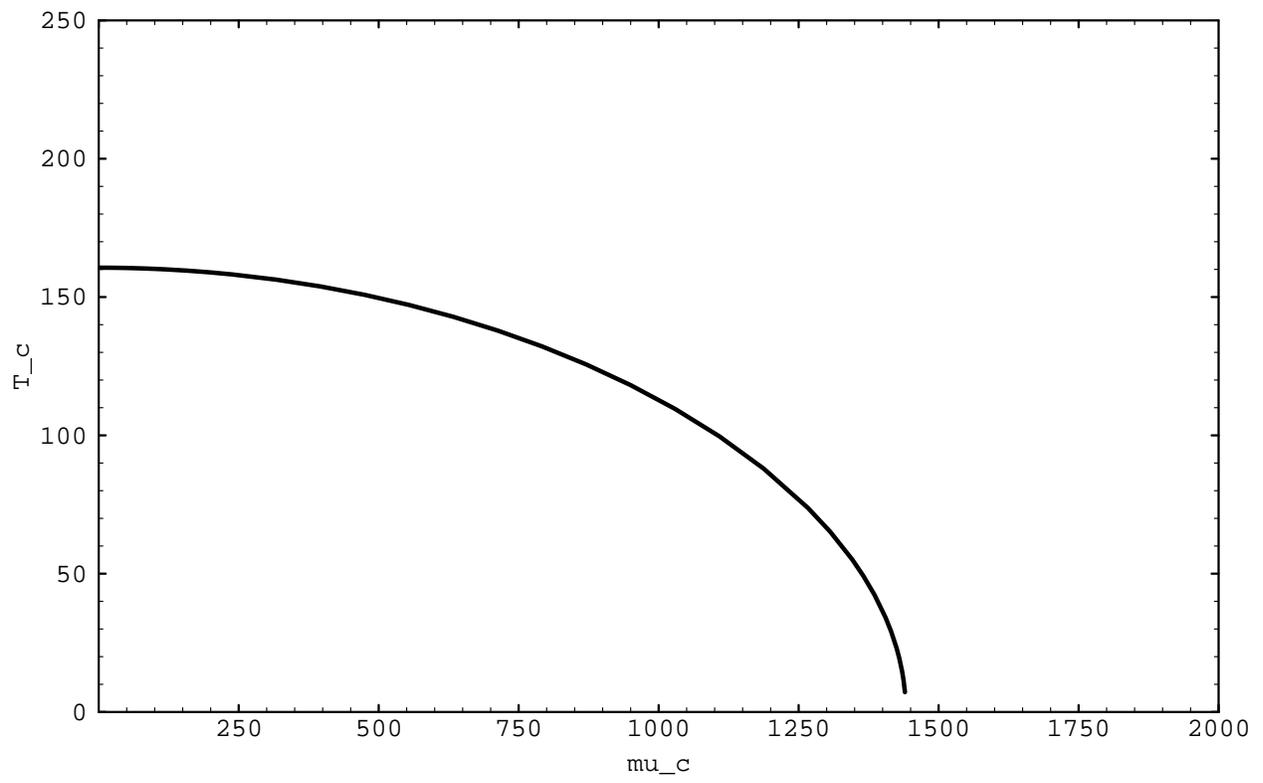
\end{document}